\documentclass[12pt a4paper]{article}

\usepackage{latexsym}
\usepackage{graphicx}
\usepackage{amsmath}
\usepackage{amssymb}
\usepackage[dvips]{color}
\graphicspath{{./}{./figs/}}

\begin{document}

\title{\bf Sync in Complex Dynamical Networks: Stability, Evolution, Control, and Application}
\author{{\bf Xiang Li\thanks{E-mail: xli@sjtu.edu.cn. This work was partly supported by the National
Natural Science Foundation of P.R. China under Grants No. 90412004
and 70431002, and partly supported by Shanghai Rising-Star Program
(No. 05QMX1436). The author also acknowledges the support
from the Alexander von Humboldt Foundation and the SRF for ROCS, SEM.}}\\
Department of Automation\\
Shanghai Jiao Tong University\\
Dongchuan Road 800, Shanghai 200240, P.R.China\\
\\(Invited Review for International Journal of Computational Cognition)\\}

\maketitle

\begin{abstract}
In the past few years, the discoveries of small-world and
scale-free properties of many natural and artificial complex
networks have stimulated significant advances in better
understanding the relationship between the topology and the
collective dynamics of complex networks. This paper reports recent
progresses in the literature of synchronization of complex
dynamical networks including stability criteria, network
synchronizability and uniform synchronous criticality in different
topologies, and the connection between control and synchronization
of complex networks as well. The economic-cycle synchronous
phenomenon in the World Trade Web, a scale-free type of social
economic networks, is
used to illustrate an application of the network synchronization mechanism.\\

{\bf keywords:} Synchronization, small-world, scale-free, pinning
control, economic-cycle\\

\end{abstract}

\section{Introduction}

Synchronization is a long-lasting fundamental concept and is, in
fact, a universal phenomenon in all areas of science and
technology
\cite{Blekhman:1988,M-M-P:2002,P-R-K:2001,Strogatz:2003,Wu:2002}.
There are many interesting synchronization phenomena seen in our
daily life, including for example fireflies flashing in unison,
crickets chirping in synchrony, and heart cells beating in rhythm.

One of the most significant and interesting properties of a
dynamical network is the synchronous output motion of its network
elements (nodes). Synchronization in coupled dynamical networks
and systems has been studied for many years within a common
framework based on nonlinear dynamical system theories, due to its
ubiquity in many technological fields such as coupled laser
systems, biochemical systems, and communication networks. In
recent years, synchronization in a network of coupled chaotic
systems has become a topic of great interest
\cite{M-M-P:2002,Wu:2002}. It has been observed, however, that
most existing work have been concentrated on networks with
completely regular topological structures such as chains, grids,
lattices, and fully-connected graphs. Two typical cases are the
discrete-time coupled map lattices \cite{Kaneko:1992} and the
continuous-time cellular nonlinear networks \cite{Chua:1998}. The
main benefit of these simple architectures is that it allows one
to focus on the complexity caused by the nonlinear dynamics of the
nodes, without considering additional complexity caused by the
network structure itself. Another appealing feature is the
possibility of realizing such regularly coupled networks by
integrated circuits, with the obvious advantage of compactness,
for potential engineering and technological applications.

The topology of a network often affects its functional behaviors.
For instance, the topology of a social network affects the
spreading of information and also disease, while an unsuitable
topology of a power grid can damage its robustness and stability.
In fact, we are confronting all kinds of networks with complex
structures everyday, where handy instances are the Internet and
the World Wide Web (WWW). Therefore, people are particularly
interested in the question of how to model such complex networks.
Traditionally, the topology of a complex network is described by a
completely random network generated by the so-called
Erd\"os-R\'enyi ({\em ER}) model \cite{E-R:1960}, which is at the
opposite end of the spectrum from a completely regular network,
and is one of the oldest and perhaps also the best tools for
study. However, with the increasing interest in trying to
understand the essence of various real-life complex networks such
as the Internet, the World Wide Web, the metabolic networks, and
various social and economic networks like the
scientific-collaboration network and the World Trade Web, etc.
\cite{A-B:2002,D-M:2002,L-J-C:2003,L-J-C:2004,
Newman:2000,S-B:2003,Strogatz:2001}), people have discovered the
so-called small-world phenomena and scale-free feature, which
cannot be well described by the ER random graph theory, and
therefore stimulate a strong desire of building new network models
\cite{A-B:2002,D-M:2002,L-C:2003-1,Newman:2000,Newman:2003}.

Watts, Strogatz and Newman (WSN), for example, introduced their
{\em small-world network models}, which generate networks having
short average path lengths along with large clusters
\cite{Newman:2000,W-S:1998,Watts:1999}. These new models show a
transition from a regular network to a random network. Both the ER
random graph and WSN small-world network models have a common
Poisson connectivity distribution and are homogenous in nature:
each node in such networks has about the same number of
connections. Another significant discovery is the {\em scale-free}
feature in a number of real-life complex networks, whose
connectivity distributions are in the power-law form as first
pointed out by Barab\'{a}si and Albert (BA)
\cite{B-A:1999,B-A-J:1999}. Owing to the non-homogenous topology,
i.e., most nodes have very few connections and only very few nodes
have many connections, scale-free networks show a unique
characteristic: robustness and yet fragility (see
\cite{A-J-B:2000,C-N-S:2000,L-W-C:2004,W-C:2002-1} for more
precise interpretations). The aforementioned small-world and
scale-free features have also been empirically verified to fit
many real-life complex networks: they are largely clustered with a
short path length, following a power-law degree distribution.

The discovery of the small-world effect and scale-free feature of
most complex networks has led to a fascinating set of common
problems concerning how a network structure facilitates and
constrains the network collective behaviors. In particular, a lot
of work have been concentrated on synchronization in small-world
and scale-free dynamical networks, including small-world networks
of oscillators
\cite{B-P:2002,H-C-K:2002,Lv-Y-C-C:2004,W-C:2002-2,Watts:1999},
small-world neural networks \cite{L-H-C-S:2000,Lu-C:2004},
small-world circle map lattices \cite{B-P-V-L:2003}, coupled map
lattices with small-world interactions \cite{G-H:2000,J-J:2002},
and scale-free networks of oscillators
\cite{L-C:2003-2,LX:2005,N-M-L-H:2003,W-C:2002-1} and maps
\cite{A-J:2003,J-A:2003,L-G-H:2004}. Not limited to the above,
synchronous phenomena in complex networks with time delays
\cite{A-J:2004,D-J-D:2004,LCG-C:2004,M-M:2005}, time-varying
coupling \cite{B-B-H:2004a,B-B-H:2004b,Lu-C:2004,Lv-C:2005,Lv-Y-C:
2004}, and weighted coupling
\cite{Chavez:2005,M-Z-K:2005-1,M-Z-K:2005-2} have been extensively
investigated, along with which many network evolving factors have
been understood to play an important role of synchrony
improvements
\cite{F-L-W:2005a,F-W:2005,H-C-K:2002,H-K-C-P:2004,M-P:2004,M-V-P:2004,N-M-L-H:2003,Wu:2003,Wu:2005}.
Of particular interest in real applications is the synchronization
phenomenon in the scale-free World Trade Web studied
\cite{L-J-C:2003}, which will be further introduced in this
article below.

Although this survey can only cover a small part of recent
progresses in this fast developed literature, we wish it could
show the audience that current wide studies of complex networks
will continuously motivate more efforts devoted to the
understanding of synchronization and many other aspects of various
complex dynamical networks and systems.

\section{Preliminaries: several models of complex network topologies}
\subsection{ER model: random networks}
\begin{figure}[h]
\centering
\includegraphics[width=10cm, height=8cm]{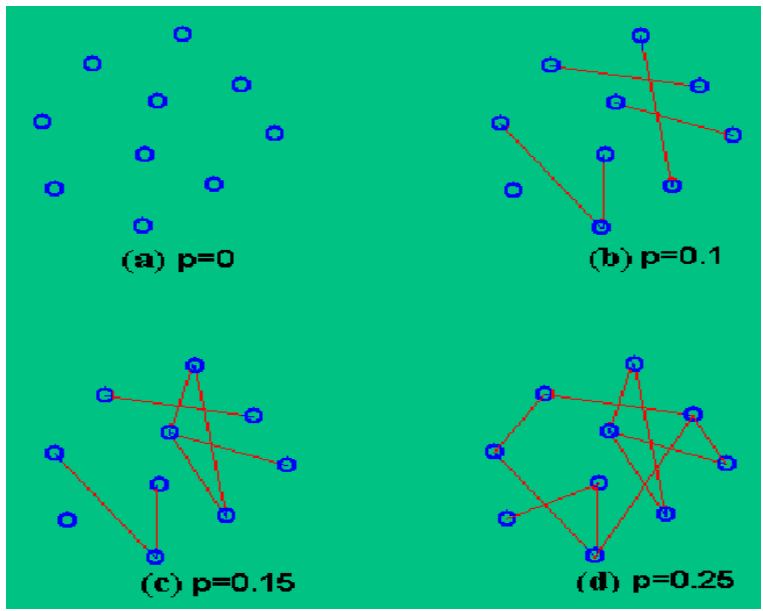}
\caption{\label{fig-ER} Evolution of an ER random network. One
starts with isolated nodes in (a), and connects every pair of
nodes with probability (b)$p=0.1$, (c) $p=0.15$, and (d) $p=0.25$,
respectively (After \cite{Wang:2002}).}
\end{figure}

The ER model of random networks \cite{E-R:1960} is defined as a
random graph (network) having $N$ labelled nodes connected by $n$
edges, which are chosen randomly from all the $\frac{N(N - 1)}{2}$
possible edges. The network evolves as follows: Start with $N$
nodes, and every pair of nodes are connected with the same
probability $p$, as shown in Figure \ref{fig-ER}.

\begin{figure}[h]
\centering
\includegraphics[width=8cm,height=6cm]{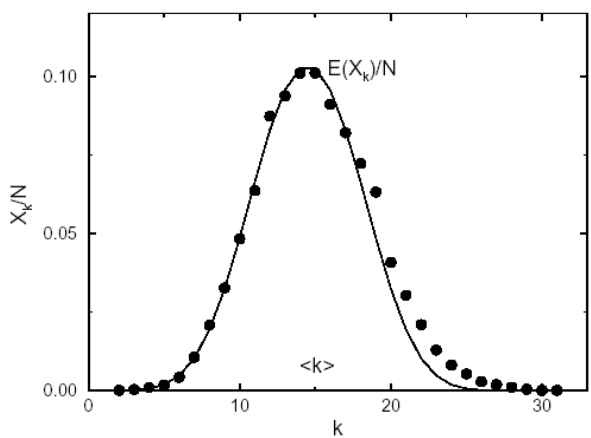}
\caption{\label{fig-HomPk} The degree distribution that results
from the numerical simulation of an ER random network, which is
generated with $N = 10,000$ and connection probability $p=0.0015$
(After \cite{A-B:2002}).}
\end{figure}

The main goal of the random graph theory is to determine, under
what connectivity probability $p$, a particular property of a
graph (network) will likely arise. For a large $N$, the ER model
generates a homogenous random network, whose connectivity
approximately follows a Poisson distribution described by (Fig.
\ref{fig-HomPk})

\begin{equation}
\label{eq-0a} P(k) \approx e^{ - \left\langle k \right\rangle
}\frac{\left\langle k \right\rangle ^k}{k!}
\end{equation}

\noindent where $\left\langle k \right\rangle $ is the average of
$k_i $, the degree of node $i$, over all nodes in the network .
With this connectivity distribution, nodes in the network are
quite uniformly spread out, which is known as a homogenous feature
of the distribution.

\subsection{WS and NW models: small-world networks}
Recall the evolving algorithm of the WS model \cite{W-S:1998}
described as follows: (I) Start with a $K-$nearest-neighbor
coupled network consisting of $N$ nodes arranged in a ring, and
each node $i$ is adjacent to its neighbor nodes $i\pm 1$, $i\pm
2$, ......, $i\pm K/2$, where $K$ is even. In order to have a
sparse but connected network, assume $1<k\ll N$ in general. (II)
Randomly rewire each edge of the network with probability $p$.
Rewiring in this context means shifting one end of the edge to a
new node chosen at random from the whole network, with the
constraint that no two different nodes can have more than one edge
in between, and no node can have an edge with itself. There are
three prominent network evolving phases in the WS small-world
model: when $p=0$, the WS evolving network stays at the original
state of a regular ring; when $p=1$, the WS network evolves to a
completely random network; only when $0<p\ll 1$, a small-world
network emerges from the WS model.

To avoid leading to the formulation of isolated clusters, which
may happen in the WS model, Newman and Watts proposed their NW
small-world model as a variant of the WS model \cite{N-W:1999},
which evolves as follows: (I) Start with a $N$-node regular ring
in which every node is connected to its first $K$ neighbors ($K/2$
on each side). Here also $1< K\ll N$. (II) Instead of rewiring in
the WS model, add a new long-range edge (short-cut) into the
network with probability $0<p\ll 1$ between randomly chosen pair
of nodes.

\begin{figure} [h]
\centering
\includegraphics[width=12cm,height=8cm]{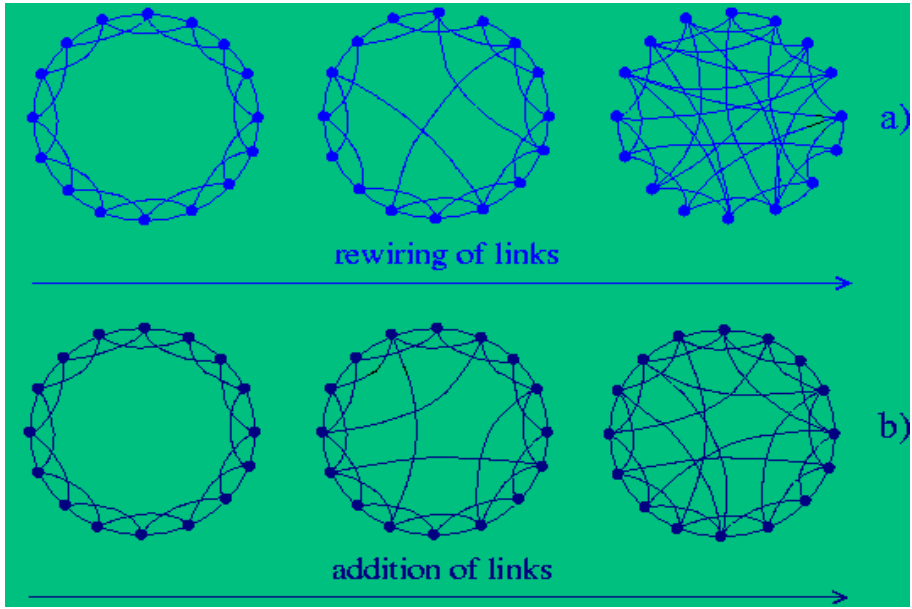}
\caption{\label{figNWS} Network evolution of small-world models:
(a) The WS model with the rewiring of edges. (b) The NW model with
the addition of edges. (After \cite{Wang:2002}).}
\end{figure}

It has been well known that only for sufficient large $N$ and very
small $p$, the NW model is then equivalent to the WS model, and
networks so-generated are hereby in the category of small-world
networks having both short average distances and large clusters
(Fig. \ref{figNWS}). Both the WS model and the NW model generate
small-world networks having the same degree distribution as that
of random networks shown in Fig \ref{fig-HomPk}.

\subsection{BA model: scale-free networks}
The algorithm of the Barab¨¢si-Albert (BA) scale-free model is in
the following \cite{B-A-J:1999}: (I) Growth: Starting with a small
number ($m_0$) of nodes, at every time step, add a new node with
$m$ ($\le m_0$) edges that link the new node to $m$ different
nodes already presented in the network. (II) Preferential
attachment: When choosing the nodes to which the new node
connects, assume that the probability $\Pi_i$ that a new node will
be connected to node $i$ depends on the degree $k_i$ of node $i$,
in such a way that
\begin{equation}
\Pi_i=\frac{k_i}{\sum_j k_j}
\end{equation}

\begin{figure} [h]
\centering
\includegraphics[width=12cm,height=9cm]{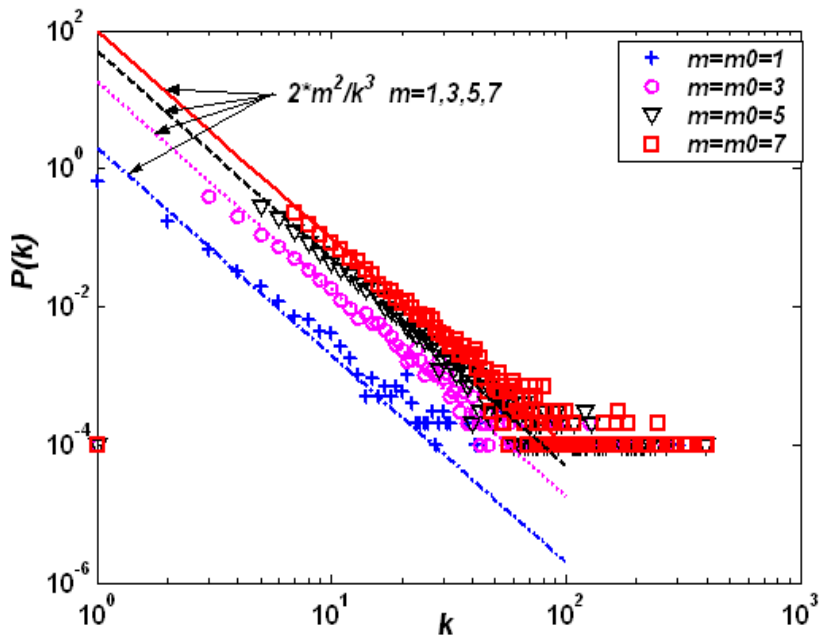}
\caption{\label{fig-BA} Degree distribution $P(k)$of scale-free
networks generated by the BA model, with $N = m_0 + t = 10000$ and
$m = m_0 = 1,3,5,7,$ respectively.}
\end{figure}

After $t$ time steps, we get a network having $m_0+t$ nodes and
$mt$ edges. This network arrives at a scale-invariant state with
the probability that a node has  $k$ edges following a power-law
distribution $P(k)\sim 2m^2k^{ - \gamma _{BA} }$ with an exponent
$\gamma_{BA}=3$ (Fig. \ref{fig-BA}), where the scaling exponent
$\gamma_{BA}$ is independent of parameter $m$ and the network
scale $N$, and in this scale-invariant sense the network is said
to be scale-free. It has been argued that the BA scale-free
network model has captured two basic mechanisms, growth and
preferential attachment, responsible for the scale-free feature
and "rich gets richer" phenomenon in many real-life and
large-scale complicated networks \cite{B-A-J:1999}.

\section{Stability criteria of dynamical network synchronization}
\subsection{The criterion of $\lambda_N/\lambda_2$}
Consider a network having $N$ identical, diffusively coupled
nonlinear oscillators as follows \cite{B-P:2002}, in which each
node is an $n$-dimensional continuous-time dynamical system:
\begin{equation}
\label{eq-1}{\dot{{\bf x}}}_{i}=f({\bf x}_{i})+c\sum_{j=1}^N
a_{ij}\,H({\bf {x_j}}), \qquad i=1,2,\cdots ,N,
\end{equation}
where ${\bf x}_{i}=[x_{i1},\, x_{i2},\ ...\,,\, x_{in}]^{T}\in
\Re^{n}$ are the state variables of node $i$, the constant $c>0$
represents the coupling strength of the network, and $H\in
\Re^{n\times n}$ is the output matrix of each node. The coupling
matrix $A=(a_{ij})\in \Re^{N\times N}$ represents the coupling
configuration of the network: if there is a connection between
node $i$ and node $j$, then $a_{ij}=1$; otherwise, $a_{ij}=0$
($i\neq j)$. Moreover, $a_{ii}=-k_{i}$.

The coupled network (\ref{eq-1}) is said to achieve ({\em
asymptotical}) {\em synchronization} if
\begin{equation}
\label{eq-2} {\bf x}_1(t)={\bf x}_2(t)=\cdots={\bf x}_N(t)\to
s(t), \mbox{ as } t\to\infty,
\end{equation}
where $s(t)\in \Re^{n}$ can be an equilibrium point, a periodic
orbit, and even a chaotic orbit, depending on the interest of
study.

Suppose that the network is connected without isolate clusters,
whose coupling matrix $A$ is therefore a symmetric irreducible
matrix. In this case, zero is an eigenvalue of $A$ with
multiplicity 1 and all the other real eigenvalues of $A$ are
strictly negative \cite{Wu-Chua:1995}, denoted by
\[
0=\lambda _{1}>\lambda _{2}\geq \lambda _{3}\geq \cdots \geq
\lambda _{N}\,.
\]

Diagonalize the variation equation of network (\ref{eq-1}), which
yields
\begin{equation}
\label{eq-2a}{\dot{\xi}}_{i}=[Df(s)+c\lambda_iDH(s)]\xi_i, \qquad
i=1,2,\cdots ,N,
\end{equation}
where $\xi_i\in \Re^{n}$ is the transversal error of node $i$ to
the synchronized states (\ref{eq-2}).

Denote $c\lambda_i=\alpha+i\beta$, where $\alpha>0$, $\beta=0$.
The variation equation (\ref{eq-2a}) comes to the so-called {\sl
Master Stability Equation} of network (\ref{eq-1}):
\begin{equation}
\label{eq-2b}{\dot{\xi}}_{i}=[Df(s)+\alpha DH(s)]\xi_i, \qquad
i=1,2,\cdots ,N,
\end{equation}
whose largest Lyapunov exponent $L_{max}$ is a function of
$\alpha$ (and $\beta=0$), the {\sl Master Stability Function}
\cite{B-P:2002,P-C:1998}.

If for every nonzero eigenvalue $\lambda_i$, $i=2,\cdots, N$, the
corresponding $L_{max}^i<0$, and the synchronized states
(\ref{eq-2}) are then stable. Having scaling transformations with
the ordered eigenvalues $\lambda_i$, $i=2,\cdots, N$, we finally
arrive at the following group of inequalities:
\begin{equation}
\label{eq-3} \left\{\begin{array}{l}
c\lambda_N>\alpha_2^2=\alpha_2\\
c\lambda_2<\alpha_1^N=\alpha_1
\end{array}
\right.
\end{equation}

Therefore, to achieve the stable synchronized states (\ref{eq-2}),
Barahona and Pecora stated that the following inequality should be
satisfied \cite{B-P:2002}:
\begin{equation}
\label{eq-4} \frac{\lambda_N}{\lambda_2}<\frac{\alpha_2}{\alpha_1}
\end{equation}

\subsection{The criterion of $\lambda_2$}
Wang, Li, and Chen considered the network (\ref{eq-1}) of $N$
identical, linearly and diffusively coupled nodes
\cite{L-C:2003-2,W-C:2002-1,W-C:2002-2}:
\begin{equation}
\label{eq-5}{\dot{{\bf x}}}_{i}=f({\bf x}_{i})+c\sum_{j=1}^N
a_{ij}\,\Gamma {\bf x}_{j}, \qquad i=1,2,\cdots ,N,
\end{equation}
where the inner linking matrix
$\Gamma=\mbox{diag}(r_1,\cdots,r_n)$ is a constant $0$-$1$
diagonal matrix linking the coupled variables, i.e., the matrix
$H$ in network (\ref{eq-1}) is a linear diagonal matrix $\Gamma$.

Wang and Chen \cite{W-C:2002-1} showed the audience the following
theorem in the sense of Lyapunov stability.

\newtheorem{theorem}{Theorem}
\begin{theorem}\label{Theorem1} Suppose there
exists an $n\times n$ diagonal matrix $P>0$ and two constants
$\overline d<0$ and $\tau>0$, such that
\begin{equation*}
[Df(s(t))+d\Gamma]^{T}P+P[Df(s(t))+d\Gamma]\le -\tau I_{n}
\end{equation*}
for all $d\le \overline{d}$, where $I_{n}\in \Re^{n}$ is a unity
matrix. If
\begin{equation}
\label{eq-6} c>\left|\frac{\overline{d}}{\lambda_2}\right|,
\end{equation}
then the synchronized states (\ref{eq-2}) of dynamical network
(\ref{eq-5}) are exponentially stable.
\end{theorem}

Further specify $\Gamma=I_n$, and assume each node (an $n$
dimensional continuous-time dynamical system) is chaotic, whose
largest Lyapunov exponent is $h_{max}>0$. The following theorem
pointed out the analytic value of constant $\overline{d}=-h_{max}$
\cite{L-C:2003-2}.

\begin{theorem}\label{Theorem2} Consider the network (\ref{eq-5}) having $N$ identical chaotic systems, where
$\Gamma=I_n$. If
\begin{equation}
\label{eq-7} c>\frac{h_{max}}{\left|\lambda_2\right|},
\end{equation}
then the synchronized states (\ref{eq-2}) of dynamical network
(\ref{eq-5}) are exponentially stable.
\end{theorem}

Two criteria of $\lambda_N/\lambda_2$ and $\lambda_2$ are not
contradictory. In fact, when we set $H=I_n$ in the variational
equation (\ref{eq-2a}), there is a bridge between the largest
Lyapunov exponent $L_{max}$ of network (\ref{eq-1}) and the
largest Lyapunov exponent $h_{max}$ of each chaotic node in the
network:
\begin{equation}
\label{eq-8} L_{max}(\lambda_i)=h_{max}+c\lambda_i, \qquad
i=1,2,\cdots,N,
\end{equation}
which finally reaches the same $\lambda_2$ criterion of Eq.
(\ref{eq-7}). Extended from these two synchronization criteria,
networks are proposed to be categorized to two classes with their
corresponding synchronizability \cite{K-A:2005}.

\subsection{The case of discrete-time coupled networks}
Consider the following discrete-time form of a complex network
with coupled maps:
\begin{equation}
\label{eq-9}{{\bf x}}_{i}(k+1)=f({\bf x}_{i}(k))+c\sum_{j=1}^N
a_{ij}\, f({\bf x}_{j}), \qquad i=1,2,\cdots ,N,
\end{equation}
where corresponding variables are defined in previous sections.
Similarly, the discrete-time coupled network (\ref{eq-9}) is said
to achieve (asymptotical) {\em synchronization} if
\begin{equation}
\label{eq-10} {\bf x}_1(k)={\bf x}_2(k)=\cdots={\bf x}_N(k)\to
s(k), \mbox{ as } k\to\infty.
\end{equation}

The counterpart of Theorem{\ref{Theorem2}} for discrete-time
coupled network (\ref{eq-9}) is as follows \cite{L-C:2003-2}.

\begin{theorem}\label{Theorem3} Consider the chaotic network (\ref{eq-9}) having $N$ identical coupled maps. If
\begin{equation}
\label{eq-11} \frac{1-e^{-h_{max}}}{\left|
\lambda_2\right|}<c<\frac{1+e^{-h_{max}}}{\left|
\lambda_N\right|}\,.
\end{equation}\\
then the synchronized states (\ref{eq-10}) of dynamical network
(\ref{eq-9}) are exponentially stable.
\end{theorem}

Therefore, condition (\ref{eq-11}) can be re-written in the form
of master stability function (\ref{eq-4}), with
$\alpha_1=1-e^{-h_{max}}$, $\alpha_2=1+e^{-h_{max}}$ in this
specified case.\\

Some latest extensions of network (\ref{eq-5}) with time-delays
\cite{LCG-C:2004,Lu-C:2004} and time-varying embedments
\cite{Lv-C:2005,Lv-Y-C: 2004} have been investigated, whose
stability conditions still belong to the $\lambda_2$ criterion as
that of the origin (\ref{eq-5}). More closely, some researchers
have noticed the synchronization stability in the framework of
graph theory \cite{A-B:2005,B-B-H:2004a,B-B-H:2004b},
 which, although not covered in this paper, deserves more
 attention and future explorations.

 \section{Synchronizability of complex evolving networks}

\subsection{Synchronizability of small-world networks}

Given the dynamics of an isolate node and the inner linking
structural matrix, the synchronizability of the dynamical network
(\ref{eq-5}), with respect to a specific coupling configuration,
is said to be {\em strong} if the network can synchronize with a
small value of the coupling strength $c$ in condition
(\ref{eq-6}).

The second-largest eigenvalue of the coupling matrix in a globally
coupled network is $-N$. This implies that for any given and fixed
nonzero coupling strength $c$, a globally coupled network will
synchronize as long as its size $N$ is large enough. On the other
hand, the second-largest eigenvalue of the coupling matrix of a
nearest-neighbor coupled network tends to zero as $N\to\infty$,
which implies that for any given and fixed nonzero coupling
strength $c$, a nearest-neighbor coupled network cannot
synchronize if its size $N$ is sufficiently large
\cite{W-C:2002-1,W-C:2002-2}.

Now, consider the dynamical network (\ref{eq-5}) with the NW
small-world connections. Let $\lambda_{2sw}$ be the second-largest
eigenvalue of the coupling matrix $A$. It was found that
\cite{W-C:2002-2}, for any given coupling strength $c$: (i) for
any $N>|\bar d|/c$, there exists a critical value $\bar p$ such
that if $\bar p\leq p\leq 1$ then the small-world network will
synchronize; (ii) for any given $p\in (0,\;1]$, there exists a
critical value $\bar N$ such that if $N>\bar N$ then the
small-world network will synchronize. These results imply that the
ability of achieving synchronization in a large-size
nearest-neighbor coupled network can be greatly enhanced by just
adding a tiny fraction of distant links, thereby making the
network a small-world. This also reveals an advantage of
small-world networks for achieving synchronization, if desired.

\subsection{Synchronizability of scale-free networks: Robust and yet
fragile}

Consider the dynamical network (\ref{eq-5}) again, but with BA
scale-free connections instead. It was found \cite{W-C:2002-1}
that the second-largest eigenvalue of the corresponding coupling
matrix is very close to $-1$, which actually is the second-largest
eigenvalue of star-shaped coupled networks. This implies that the
synchronizability of a scale-free network is about the same as
that of a star-shaped coupled network. It may be due to the
extremely inhomogeneous connectivity distribution of such
networks: a few ``hubs'' in a scale-free network play a similar
important role as a single center in a star-shaped coupled
network.

The robustness of synchronization in a scale-free dynamical
network has also been investigated \cite{W-C:2002-1}, against
either random or specific removal of a small fraction $\delta$ of
nodes from the network. Obviously, the removal of some nodes from
network (\ref{eq-5}) will change the coupling matrix. If the
second-largest eigenvalue of the coupling matrix remains
unchanged, then the synchronization stability of the network will
remain unchanged after the removal of such a node. Let $A\in
\Re^{N\times N}$ and $\tilde{A}\in \Re^{N\times N}$ be the
coupling matrices of the original network with $N$ vertices and
the new network after removal of $[\delta N]$ vertices,
respectively. Denote $\lambda_2$ and $\tilde{\lambda}_2$ as the
second-largest eigenvalues of $A$ and $\tilde{A}$, respectively.

It was found \cite{W-C:2002-1} that even when as many as $5\%$ of
randomly chosen nodes are removed, the second-largest eigenvalue
of the coupling matrix, $\tilde{\lambda}_2$, remains almost
unchanged; therefore, the ongoing synchronization is not altered.
On the other hand, although the scale-free structure is
particularly well-suited to tolerate random errors, it is also
particularly vulnerable to deliberate attacks. In particular,
under an intentional attack, the magnitude of $\tilde{\lambda}_2$
decreases rapidly; for example, it almost decreases to one half of
its original value of $\lambda_2$ in magnitude, when only $1\%$
fraction of the highly connected nodes were removed. Therefore, it
is very reasonable to believe that the error tolerance and attack
vulnerability of synchronizability in scale-free networks are
rooted in their extremely inhomogeneous connectivity patterns.

\subsection{Synchronizability preference and network robustness}
However, the preference attachment in the BA model of scale-free
networks may not directly result in its fragility to specific
removal of a small fraction of (highly connected) nodes. We make
further exploring with the introduction of synchronous
preferential attachment mechanism into the BA model, and the
probability $\Pi_i$ with which the new node is connected to node
$i$ depends on the synchronizability (characterized by
$\lambda_{2i}$) of the new network if the new node had connected
to node $i$ , such that \cite{F-L-W:2005a}
\begin{equation}
\label{eq-12} \Pi_i=\frac{\lambda_{2i}}{\sum_j \lambda_{2j}}.
\end{equation}
After $t\ll m_0$  time steps, we obtain a {\sl
synchronization-preferential} network with $m_0+t$ nodes and $mt$
edges.

\begin{figure} [t]
\centering
\includegraphics[width=10cm]{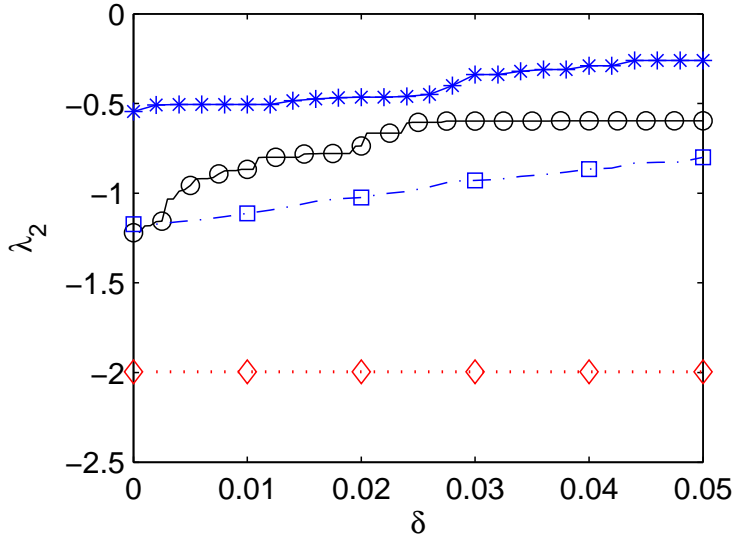}
\caption{\label{fig-rob} Network synchronization robustness
against random failures: changes of $\lambda_2$ of the BA
scale-free networks (solid line with circles), the
synchronization-optimal networks (dotted line with diamonds), the
ER random networks (solid line with stars), and the
synchronization-preferential networks (dash-dotted line with
squares) when a fraction $\delta$ of randomly selected nodes are
removed. All curves are averaged from 5 groups of networks in
simulations.}
\end{figure}

An extreme case is that instead of the preferential attachment
(\ref{eq-12}), when a new node is added to the network, the
criterion for choosing the  $m$ nodes to which the new node
connects is to maximize the synchronizability of the obtained
network, or equivalently, to minimize the second-largest
eigenvalue of the corresponding coupling matrix, which comes to a
{\sl synchronization-optimal} network \cite{F-W:2005}.

Now we consider the robustness of synchronization in dynamical
network (\ref{eq-5}) against either random or specific removal of
a small fraction $\delta$ of vertices in these newly proposed
networks. In extensive numerical simulations, it has been shown
that for the synchronizability of a synchronization-preferential
network, its error tolerance to random removal is a little better
than that of a BA scale-free network. The value of $\lambda_2$
decreased from -1.2493 to -0.800 when 5\% of the vertices were
randomly removed, which has none significant reduction when more
nodes were randomly removed. This implies that the proposed
synchronization-preferential model is robust against random
failures.

Move to the synchronization fragility of the
synchronization-preferential network with respect to deliberate
attacks. At every time step, we remove the nodes with the highest
degree and found that the second-largest eigenvalue of coupling
matrix decreased from -1.2493 to -0.3802 when as many as 5\% of
the most connected nodes were removed on purpose. Furthermore, the
decreased magnitude of $\lambda_2$ of a
synchronization-preferential network is smaller than that of a BA
network. Therefore, a synchronization-preferential network is less
vulnerable to specific removal of those most connected nodes than
a BA scale-free network.

\begin{figure} [h]
\centering
\includegraphics[width=10cm]{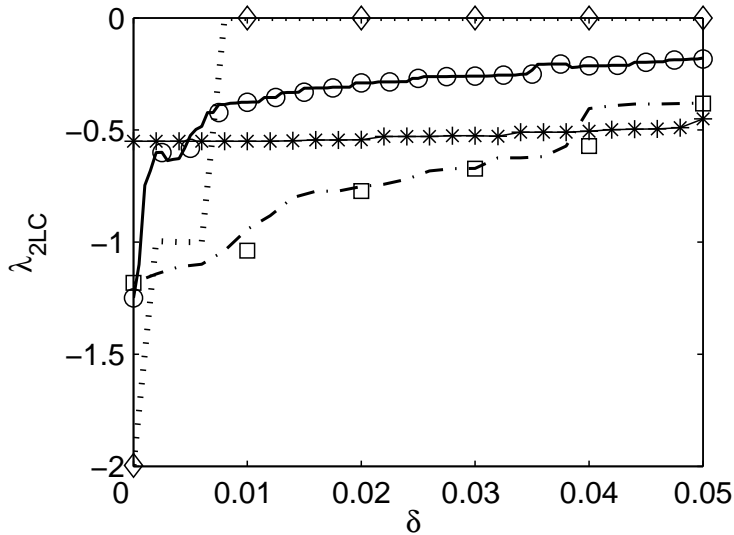}
\caption{\label{fig-frag} Network synchronization fragility
against specific attacks: changes of $\lambda_2$ of the largest
cluster in the BA scale-free networks (solid line with circles),
the synchronization-optimal networks (dotted line with diamonds),
the ER random networks (solid line with stars), and the
synchronization-preferential networks (dash-dotted line with
squares), when a fraction $\delta$ of the most connected nodes are
purposely removed.  All curves are averaged from 5 groups of
networks in simulations.}
\end{figure}

These conclusions could be visualized through the comparisons of
the robustness and fragility of synchronizability among the ER
model, the BA model, the synchronization-optimal model, and the
synchronization-preferential model (Figs.
\ref{fig-rob}-\ref{fig-frag}). The illustrations imply that the
preferential attachment mechanism does not necessarily lead to the
robust-and-yet-fragile ``Achilles heel'' of (BA) scale-free
networks.

\subsection{Synchronizability of growing networks}
The existence of lower bound and upper bound of the coupling
strength $c$ in inequality (\ref{eq-11}) for synchronizing
discrete-time networks is similar to the master stability
synchronization in continuous-time network. Here, through the
instance of discrete-time dynamical networks, we investigate the
effect of growing mechanism on the network synchronization.

Define $R:=\frac{\lambda_1-\lambda_2}{\lambda_2-\lambda_N}$, which
measures the distance from the first eigenvalue to the main part
of the spectral density $\rho(\lambda)$ normalized by the
extension of the main part \cite{F-D-B-V:2001}. Therefore, the
invalidity of condition (\ref{eq-11}) equals to the following
inequality is satisfied
\begin{equation}
\frac{1}{R}<\frac{2e^{-h_{max}}}{1-e^{-h_{max}}}
\end{equation}
i.e., to synchronize such a coupled dynamical network of maps, the
connection density of the network should satisfy the bound
determined by its individual nodes.

\begin{figure} [h]
\centering
\includegraphics[width=10cm]{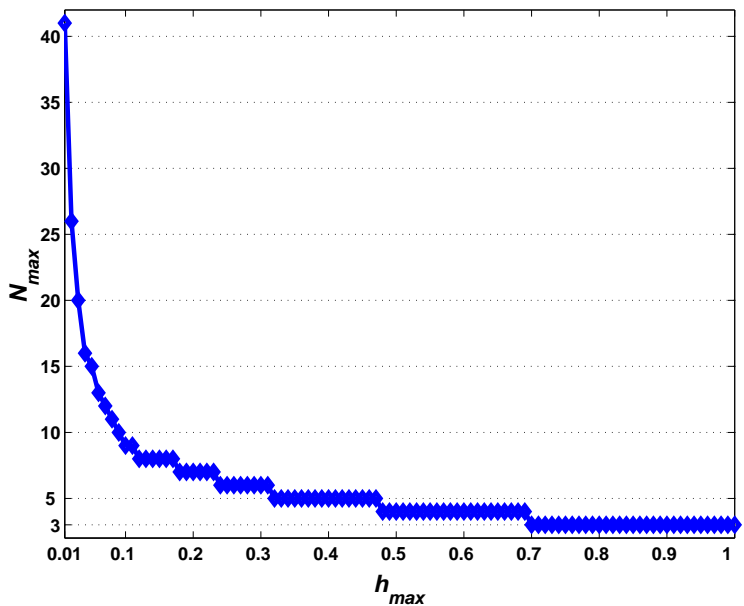}
\caption{\label{fig2} Distribution of the maximum scale-free
network size $N_{max}$ versus $h_{max}$ in [0.1, 1.0]\ (averaged
output of 20 random groups) by the BA model with $m=m_0=1$.}
\end{figure}

\begin{figure} [h]
\centering
\includegraphics[width=10cm]{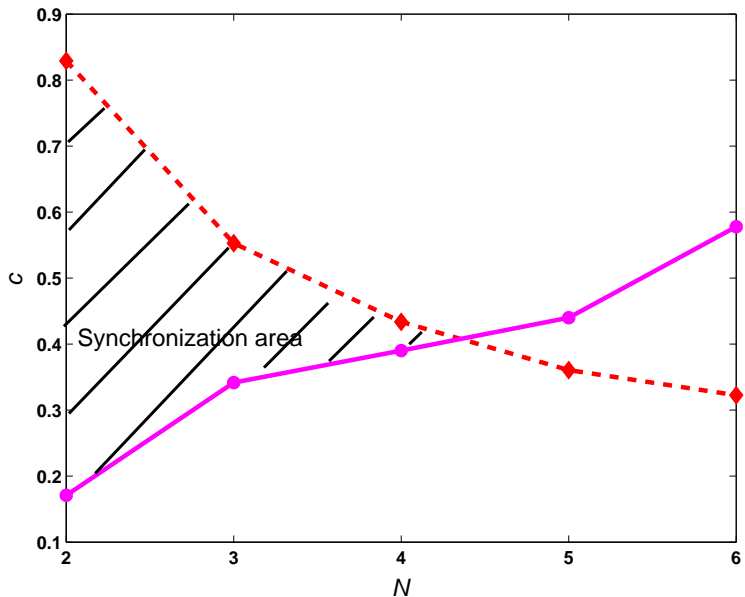}
\caption{\label{fig3} Synchronization area of the network size for
a scale-free network of coupled H\'{e}non maps (averaged output of
20 random groups). The dotted line is
$\frac{1+e^{-h_{max}}}{\left| \lambda_N\right|}$, and the solid
line is $\frac{1-e^{-h_{max}}}{\left| \lambda_2 \right|}$.}
\end{figure}

As an example of the basic BA model for scale-free networks, with
the growing mechanism of adding new nodes one by one, there exists
a maximum synchronous network scale $N_{max}$ (Fig. \ref{fig2}). A
network generated by the BA model will not synchronize if the
network further expands to have more than $N_{max}$ nodes, because
the condition (\ref{eq-11}) is then no longer satisfied
\cite{L-C:2003-2}. For example, if each node is a H\'{e}non map,
\begin{equation}
\label{Eq-13}y(k)=-1.4y(k-1)^2+0.3y(k-2)+1.0\,,
\end{equation}
the maximum network scale $N_{max}=4$ for the positive Lyapunov
exponent of the map: $h_{max}=0.418$ (Fig. \ref{fig3}).

Therefore, there are two conditions to be satisfied for
synchronizing a discrete-time dynamical network: the network scale
$N\le N_{max}$ and the inequality (\ref{eq-11}) \cite{L-C:2003-2}.
For example, as stated before, the value $N_{max}$ of a network of
coupled H\'{e}non maps for achieving synchronization is 4. Given a
4-node network of coupled H\'{e}non maps in a star-shaped
structure, condition (\ref{eq-11}) is further changed to
\begin{equation*}
\frac{1-e^{-h_{max}}}{\left|
\lambda_2\right|}=0.3416<c<\frac{1+e^{-h_{max}}}{\left|
\lambda_N\right|}=0.4146\,.
\end{equation*}
So, if the coupling strength $c$ is less than 0.3416 or larger
than 0.4146, this 4-node star of coupled H\'{e}non maps cannot
synchronize although its network scale $N\le N_{max}=4$. However,
if the coupling strength is suitable for synchronization, i.e.,
$0.3416<c<0.4146$, but a new H\'{e}non node is added to the
synchronized 4-node star, thereby obtaining a 5-node network
(thus, $N>N_{max}=4$), which cannot achieves synchronization
finally.

\section{From synchronization to control}
Now we want to stabilize network (\ref{eq-5}) onto a homogeneous
stationary state defined by
\begin{equation}
\label{eq-15} {\bf x}_1 = {\bf x}_2 = \cdots = {\bf x}_N = \bar
{\bf x}, \quad f(\bar {\bf x}) = 0.
\end{equation}

To achieve such a goal (\ref{eq-15}), we apply feedback
controllers on a small fraction $\delta (0 < \delta \ll 1)$ of the
nodes in network (\ref{eq-5}). Suppose that nodes $i_1 ,i_2
,\ldots ,i_l $ are selected, where $l = \left\lfloor {\delta N}
\right\rfloor $ stands for the smaller but nearest integer to the
real number $\delta N$. This controlled network can be described
in the re-ordered form:

\begin{equation}
\label{eq-16}\left \{
\begin{array}{l}
\dot {\bf x}_i = f({\bf x}_i ) + c\sum\limits_{j = 1}^N a_{ij}
\Gamma {\bf x}_j - cd\Gamma ({\bf x}_i - \bar {\bf x}), \quad i = 1,2, \cdots ,l \\
\dot {\bf x}_i = f({\bf x}_i ) + c\sum\limits_{j = 1}^N a_{ij}
\Gamma {\bf x}_j ,\quad\quad\quad\quad i = l + 1,l + 2, \cdots ,N
\end{array} \right.
 \end{equation}
where feedback control gain $d>0$. Generally, the number of
controllers is preferred to be very small compared with the whole
network scale $N$, i.e., $l\ll N$, and the feedback control method
in controlled network (\ref{eq-16}) is the so-called pinning
control \cite{L-W-C:2004,W-C:2002-3}. Through this section, we
mainly show a bridge over the pinning control of complex networks
and the network synchronization \cite{L-W-C:2004}.

Owing to the local error-feedback nature of each pinned node, it
is guaranteed that, as the feedback control gain $d \to \infty $,
the states of the $l$ controlled nodes can be pinned to the
homogenous target state $\bar {\bf x}$. Hence, the pinning control
stability of network (\ref{eq-16}) is converted to that of the
following virtually controlled network, and those pinned nodes
function as virtual controllers to those un-pinned nodes:

\begin{equation}
\label{eq-17}\left \{
\begin{array}{l}
{\bf x}_i = \bar {\bf x},{\begin{array}{*{20}c}
 \hfill & {i = 1,2,\ldots ,l} \hfill \\
\end{array} } \\
 \dot {\bf x}_i = f({\bf x}_i ) + c\sum\limits_{j = l + 1}^N {\tilde {b}_{ij} \Gamma
{\bf x}_j } + \tilde {u}_i ,{\begin{array}{*{20}c}
 \hfill & {i = l + 1,l + 2,\ldots ,N} \hfill \\
\end{array} } \\
 \end{array}\right.
\end{equation}
where the virtual control laws are taken as
\begin{equation}
\label{eq-18} \tilde {u}_i = - c\tilde {d}_i \left( {{\bf x}_i -
\bar {\bf x}} \right),{\begin{array}{*{20}c}
 \hfill & {i = l + 1,l + 2,\ldots ,N} \hfill \\
\end{array} }
\end{equation}
with the virtual control feedback gains $\tilde {d}_i =
\sum\limits_{j = 1}^l {a_{ij} } $, and $\tilde {B} = \left(
{\tilde {b}_{ij} } \right) \in R^{(N - l)\times (N - l)}$ defined
as

\begin{equation}
 \left\{ {\begin{array}{l}
 \tilde {b}_{ij} = a_{ij} ,{\begin{array}{*{20}c}
 \hfill & {j \ne i,j = l + 1,\ldots ,N,} \hfill \\
\end{array} }i = l + 1,l + 2,\ldots ,N \\
 \tilde {b}_{ii} = - \sum\limits_{j = l + 1,j \ne i}^N {a_{ij} }
{\begin{array}{*{20}c}
 \hfill & \hfill \\
\end{array} } \\
 \end{array}} \right.
\end{equation}

\begin{figure} [h]
\centering
\includegraphics[width=12cm]{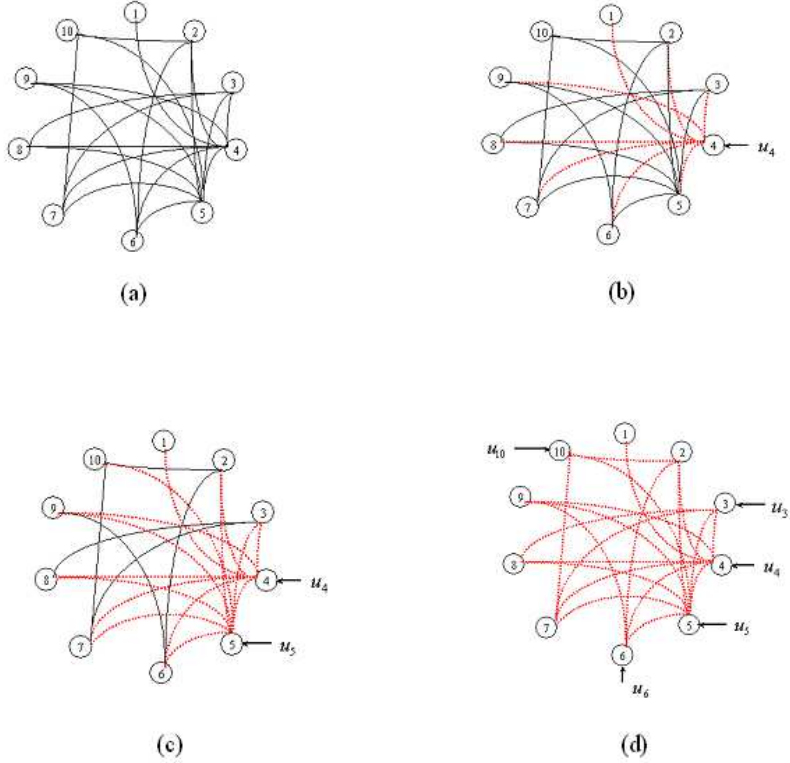}
\caption{\label{fig-control}An illustration of virtual control
(dotted line) in specifically pinning control of a network
generated by the BA model with $N=10$, $m=m_0=3$.}
\end{figure}

Figure \ref{fig-control} is an illustration of the virtual control
process in pinning controlling a network generated by the BA model
with $N = 10,m = m_0 = 3$. It can be observed that while the
pinned nodes are stabilized onto the homogenous state $\bar {\bf
x}$, their dynamics spread in the network as shown by the (red)
dotted lines, which functions as the virtual control to those
nodes which were not placed controllers. At the same time, it can
be found in Fig. \ref{fig-control}(b) that the whole 10-node
original network was broken into two parts: An isolated 1$^{st}$
node, which is only connected with the 4$^{th}$ node, and the rest
8-node subnetwork. While increasing the pinning fraction of nodes,
this subnetwork continues to shrink and finally reaches the case
as shown in Fig. \ref{fig-control}(d): every unpinned node is
`isolated' and virtually controlled by other pinned and also
stabilized nodes.

Denote
\[ \tilde {A} = \tilde {B} + diag\left( {\tilde {d}_{l + 1}
,\tilde {d}_{l + 2} , \cdots ,\tilde {d}_N } \right),
\]
and set $\Gamma=I_n$. We have the following theorem on the pinning
controlled network (\ref{eq-17}) \cite{L-W-C:2004}.

\begin{theorem}
\label{Theorem4} Assume $f(x)$ is Lipschitz continuous in $x$ with
a Lipshitz constant $L_c^f>0$, and the chaotic node $\dot {\bf
x}_i = f({\bf x}_i )$, for all $i = 1,2, \cdots ,N,$ has the
largest positive Lyapunov exponent $h_{\max}>0$.If $\tilde{A}$ is
irreducible, then the dynamical network (\ref{eq-17}) with virtual
controllers (\ref{eq-18}) is globally (or locally asymptotically)
stable about homogenous state $\bar {\bf x}$, provided that
$\Gamma = I_n $ and

\begin{equation}
\label{eq-19} c > \frac{Q}{\lambda _{\min } ( - \tilde {A})}
\end{equation}

\noindent with $Q$ being the Lipschitz constant $L_c^f$ (or the
largest Lyapunov exponent $h_{max}$).
\end{theorem}

It is very easy to verify from Theorem \ref{Theorem2} that
condition (\ref{eq-19}) is exactly the same stable condition for
synchronizing network (\ref{eq-20}):
\begin{equation}
\label{eq-20} \dot {\bf x}_i = f({\bf x}_i ) + c\sum\limits_{j = l
+ 1}^N {a_{ij} \Gamma {\bf x}_j } ,{\begin{array}{*{20}c}
 \hfill & {i = l + 1,l + 2,\ldots ,N.} \hfill \\
\end{array} }
\end{equation}

\begin{figure} [h]
\centering
\includegraphics[width=12cm]{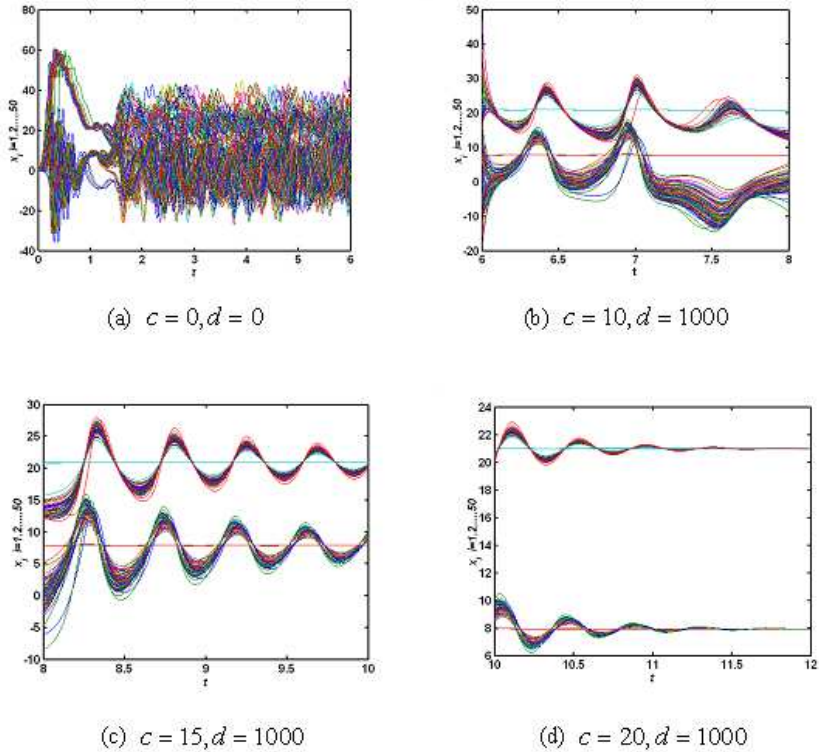}
\caption{\label{fig-contrsyn} Specifically pinning the biggest
node of 19 degrees in a 50-node Chen network generated by the BA
model: (a)-(d) are stabilizing phases with different coupling
strengths. The homogenous stationary state is $\bar{\bf
x}=[7.9373\quad 7.9373\quad 21]^T$, an unstable fixed point of
Chen system \cite{C-U:1999}.}
\end{figure}

Hence, the whole pinned network, controlled by a very small amount
of local feedback controllers and those virtual controllers, is
stabilized onto the homogenous state $\bar{\bf x}$ via
synchronization \cite{L-W-C:2004}. This phenomenon is illustrated
in Fig. \ref{fig-contrsyn} when the biggest node in a 50-node Chen
network generated from the BA model was specially pinned with
suitable coupling strength $c$ and feedback gain $d$, over which
other 49 coupled chaotic Chen systems \cite{C-U:1999} (i.e.,
nodes) in the network were synchronized to the designer-preferred
unstable fixed point $\bar{\bf x}=[7.9373\quad 7.9373\quad 21]^T$
of Chen system.

\section{Sync criticality stories on Kuramoto models}
Collective phase synchronization in a large population of
oscillators having natural different frequencies is another main
branch in the literature of network synchronization, and the
Kuramoto model is one of the most representative models of coupled
phase oscillators. After Wiener first described the mathematical
connection between the nonlinear dynamics and statical physics
\cite{Wiener:1961}, which was later fruitfully studied by Winfree
\cite{Winfree:1967}, Kuramoto refined the model, and formalized
the solution to a network of globally coupled limit-cycle
oscillators \cite{Kuramoto:1984}, uncovering the situation why the
oscillators are completely de-synchronized until the coupling
strength overcomes a criticality $C_{syn}$. Many variations and
extensions of the original Kuramoto model appeared later have been
recently surveyed in \cite{Acebron:2005} (and references therein).

We consider a network of $N$ coupled limit-cycle oscillators whose
phases $\theta_i, i=1,2,\cdots,N,$ evolve as
\begin{equation}
\label{eq-21}\frac{d\theta_i}{dt}=\omega_i+\sum_{j=1}^N
c_{ij}a_{ij}\mbox{sin}(\theta_j-\theta_i),
\end{equation}
where $c_{ij}$ is the coupling strength between node (oscillator)
$i$ and node (oscillator) $j$. Frequencies $\omega_i,
i=1,2,\cdots, N,$ are randomly distributed following the given
frequency distribution $g(\omega)$, which is assumed that
$g(\omega)=g(-\omega)$. The network coupling matrix $A=(a_{ij})\in
\Re^{N\times N}$ is defined as in previous sections.

If we select the identical coupling scheme for a globally coupled
network as
\begin{equation}
c_{ij}=\frac{c}{N}
\end{equation}
Eq. (\ref{eq-21}) is the classical Kuramoto model
\cite{Kuramoto:1984}, whose synchronization criticality is
$C_{syn}=\frac{2}{\pi g(0)}$. With similar constant identical
coupling schemes, Kuramoto model has been investigated its
criticality $C_{syn}$ over small-world and (or) scale-free
networks. Hong {\sl et al.} reported their synchronization
observations of a larger criticality of coupling strength on
small-world networks than that of globally coupled networks
\cite{H-C-K:2002}. And, the latest investigation stated the
absence of critical coupling strength in frequency synchronization
of a swarm of oscillators connected as a scale-free network having
a power-law exponent $2<\gamma\le 3$ \cite{Ichinomiya:2004}. These
results strongly suggest that the collective synchronous behaviors
of complex networks depend on the network topologies.

However, the situation is completely different when the coupling
scheme is non-identical and asymmetric. Define the non-identical
asymmetric coupling scheme \cite{LX:2005}:
\begin{equation}
\label{eq-22} c_{ij}=c_j=\frac{c}{k_i},\qquad i,j=1,\cdots,N,
\end{equation}
where $k_i$ fits the given degree distribution $P(k)$ of a
network. Therefore, we have
\begin{equation}
\label{eq-23}\frac{d\theta_i}{dt}=\omega_i+\frac{c}{k_i}\sum_{j=1}^N
a_{ij}\mbox{sin}(\theta_j-\theta_i).
\end{equation}
If for every node $i$, its degree $k_i=N, i=1,2,\cdots, N$, model
(\ref{eq-23}) is degraded to the classical Kuramoto model for
globally coupled networks. An interesting finding \cite{LX:2005}
is that, with the non-identical and asymmetric coupling scheme
(\ref{eq-22}), there does exist a uniform criticality of coupling
strength, $C_{syn}=\frac{2}{\pi g(0)}$, to synchronize diversely
random networks of limit-cycle oscillators (\ref{eq-23}), which is
surprisingly the same as the critical coupling strength of
globally coupled networks.

\begin{figure}[h]
\centering
\includegraphics[width=10cm]{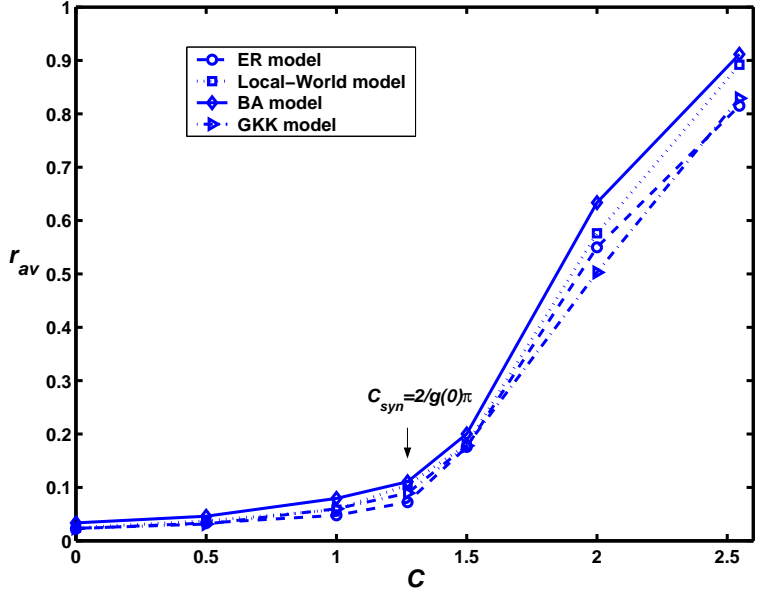}
\caption{\label{fig-KuraNet2048Rav} The average order parameter
$r_{av}$ vs the coupling strength $C$ for networks generated by
the ER model (dashed line with circle markers), the Local-World
model (dotted line with square markers), the BA model (solid line
with diamond markers), and the GKK model having power-law exponent
$\gamma=6$ (dash-dot line with right triangle markers). All
networks have the same scale $N=2048$ and the same average degree
$\left<k\right>=6$.}
\end{figure}

Two main categories of degree distributions of random complex
networks are in the forms of $P_{power}(k)\propto k^{-\gamma}$ and
$P_{exp}(k)\propto e^{-k}$. Therefore, to verify the independence
of $C_{syn}$ on different degree distributions $P(k)$ of network
topologies, in numerical studies we select the famous ER model
\cite{E-R:1960} to generate networks having an exponential degree
distribution $P_{exp}(k)\propto e^{-k}$, and the BA model
\cite{B-A-J:1999} and the GKK model \cite{G-K-K:2001} to generate
networks having the scale-invariant power-law degree distribution
$P_{power}(k)\propto k^{-\gamma}$ with $\gamma=3,6$, respectively.
The proposed local-world evolving model
 \cite{L-C:2003-1}, which owns a transition between the exponential degree distribution and the
power-law degree distribution, is adopted as the final prototype
with the parameters $M=10$, $m=m_0=3$  \cite{LX:2005}.

We fix the network scale $N=2048$, and the average degree
$\left<k\right>=6$. Therefore, all networks have the same number
of nodes and edges with different connectivity patterns in the
following simulation studies. Define the average order parameter
as
\begin{equation}
r_{av}=\left<\left[\frac{\sum_{i=1}^Nk_ie^{i\theta_i}}{\sum_{i=1}^N
k_i}\right]\right>,
\end{equation}
where $\left<\cdots\right>$ and $[\cdots]$ denote the averages
over different realizations of intrinsic networks and over
different realizations of intrinsic frequencies, respectively. In
all the simulations, $r_{av}$ is averaged over 10 groups of
networks satisfying the same degree distribution $P(k)$, and each
network has 5 sets of frequencies with the distribution
$g(\omega)$. We further specify
\begin{equation}
g(\omega)=\left\{\begin{array}{l}
0.5,\qquad\mbox{if} -1<\omega<1\\
0,\qquad\quad\mbox{otherwise}.\end{array}\right .
\end{equation}
Therefore, $C_{syn}\approx 1.273$ in this case.

As shown in Fig. \ref{fig-KuraNet2048Rav}, there is a common
critical coupling strength of the value
$C_{syn}=\frac{2}{g(0)\pi}$ in these four categories of networked
limit-cycle oscillators, and the average order parameter $r_{av}$
increased sharply when $C_{syn}\le c$. In other words, the
significant difference among complex network topologies does not
show effect on the collective synchrony of the non-identically
asymmetrically coupled limit-cycle oscillators (\ref{eq-23}).
However, as concluded from
\cite{H-C-K:2002,M-V-P:2004,Ichinomiya:2004}, there is no such a
same critical coupling strength of collective synchrony in
different random complex networks of identically symmetrically
coupled limit-cycle oscillators.

\section{An application example: Synchronization in the World Trade Web}
\begin{figure} [h]
\centering
\includegraphics[width=10cm]{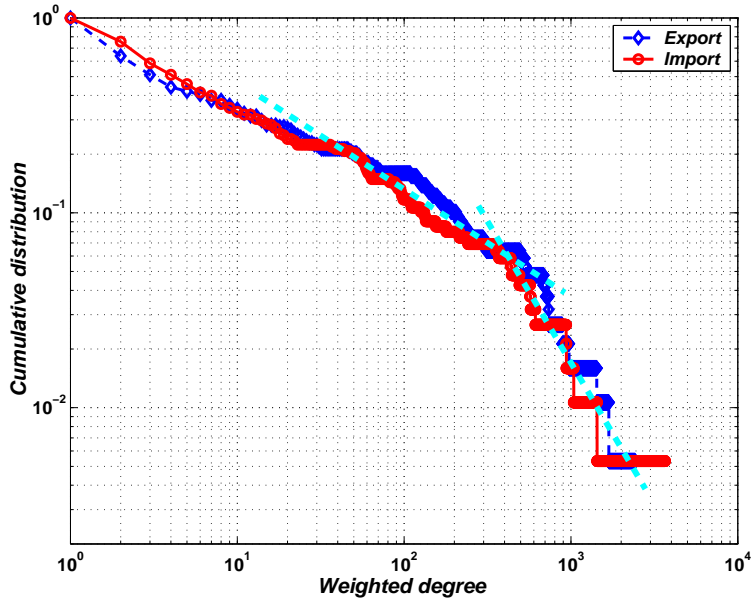}
\caption{\label{fig4}  Cumulative import and export weighted
degree distributions of the WTW with 188 nodes and 12413 export
links and 12669 import links. The dashed line is the power-law
form $P(k)\sim k^{-\gamma}$ with $\gamma=0.6\sim 1.6$.}
\end{figure}

Recent studies of the World Trade Web (WTW)
\cite{L-J-C:2003,S-B:2003} have shown that there is also a
significant scale-free feature in this economic network (Fig.
\ref{fig4}). In such a World Trade Web, every country is a
dynamical node, and the connections (edges) between any pair of
countries (nodes) are their imports and exports. It was pointed
out that the United States is the biggest node in the WTW
\cite{L-J-C:2003}. As mentioned above, the complexity of the
network topology usually dominates the dynamical behaviors of a
network, and the stability of a few `big' nodes determine the
synchronizability and stability of a scale-free dynamical network
\cite{L-W-C:2004, W-C:2002-1}. Therefore, it is interesting to ask
to what extent the economy of the United States affects the
economic development in other relatively `smaller' countries. In
more subtle details, we are interested in finding whether there is
synchronization of economic cycles existing between the United
States and the other countries.

The economic-cycle synchronization is characterized by the
correlation between the cyclical components of outputs of two
arbitrarily chosen countries, $i$ and $j$
\cite{F-R:1997,F-R:1998}:
\begin{equation}
\label{Eq-14} corr(y_i^c,y_j^c)=\frac{cov(y_i^c,y_j^c)}
{\sqrt{var(y_i^c)var(y_j^c)}}, \qquad i,j=1,2,\cdots,n\,,
\end{equation}
where $y_i^c$ is the cyclical component of output of country $i$,
which can be specified here as the real Gross Domestic Product
(GDP). A positive correlation means that synchronization exists in
the economic cycles between countries $i$ and $j$, and a higher
correlation implies a higher degree of synchronization of economic
cycles between two countries.

\begin{figure} [h]
\centering
\includegraphics[width=10cm]{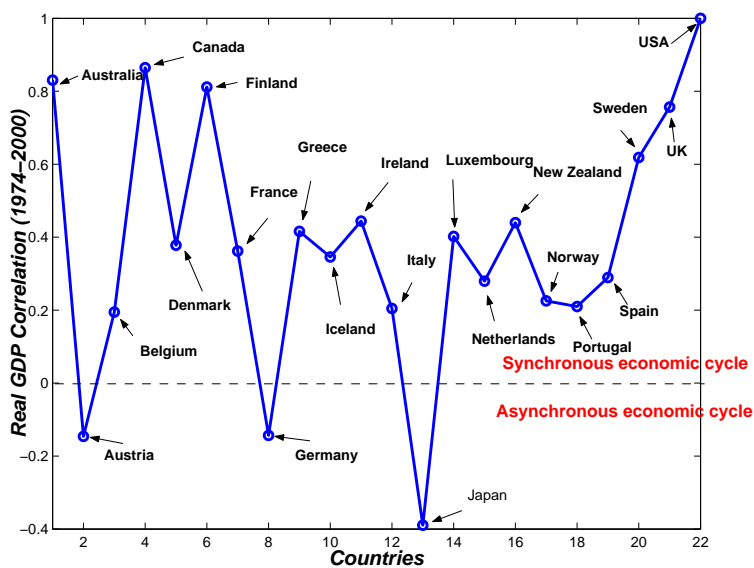}
\caption{\label{fig5}  22 developed countries' economic cycles
synchronization phenomena. The positive real GDP correlation means
synchronous economic cycles of those indicated countries with the
United States, and the negative correlation means asynchronous
economic cycles of the indicated countries with the United
States.}
\end{figure}

The economy development in a country is a very complex issue, and
may be affected by many unpredictable factors such as significant
changes of economic or political regimes occurred in a country
(like USSR and some eastern European countries), the shocks from
those speculated money firms (such as the Quantum Fund of George
Soros), the continuation of a country's economic policy, the
change of exchange rate policy, and local wars, etc. Hence, in
studying the synchronization phenomenon of economic cycles, to
decrease the perturbations on economy development from
unreasonable `noise' to the minimum, we should probe those
countries where stable and peaceful political situations and
mature market economic systems have been well maintained. With the
statistical data of real GDP in 1975-2000 that we can collected at
that time, an appropriate choice contains 22 developed countries,
i.e., the United States, the United Kingdom, Germany, Japan,
France, Canada, Australia, Austria, Belgium, Norway, Italy,
Finland, Denmark, Greece, Ireland, Iceland, Netherlands, Portugal,
Spain, Sweden, New Zealand and Luxemburg (Fig. \ref{fig5}). It can
be observed from the figure that in 1975-2000, totally 18
developed countries did have significant economic-cycle
synchronization with the United States, where 5 countries (the
United Kingdom, Australia, Canada, Finland, Sweden) show much
stronger synchronization of economic cycles, except only 3
countries: Japan, Germany and Austria \cite{L-J-C:2003}.

\section{To probe further}

In the past few years, advances in complex dynamical networks have
uncovered some amazing similarities among many diverse large-scale
natural and artificial systems such as the Internet, the world
wide web, the world trade web, cellular nonlinear networks,
metabolic systems, and even the collaborations of the Hollywood
movie stars. Researchers from inter-disciplines have been inspired
by the newly discovered small-world and scale-free features, and
significant progress has been gained in the studies of the effect
of complex network topology on network dynamical behaviors,
particularly the network synchronization phenomenon.

In this article, we have reported our recent research work on the
synchronization of complex dynamical networks, both from the
theoretical studies and economic applications. Even with the
inspiring advances of today, there are still many important
theoretical and technical explorations on modelling, evolution,
analysis, control, and synchronization of complex dynamical
networks left for future persistent research.

\section*{Acknowledgement}
The author thanks X.F. Wang and G. Chen for their valuable
contributions in the development of this paper.

\end{document}